\documentclass[prd,superscriptaddress,amsfonts,amssymb,amsmath,showpacs]{revtex4-2}
\usepackage{bm}
\usepackage{amsfonts}
\usepackage{latexsym}
\usepackage[latin1]{inputenc}
\usepackage{graphicx}
\usepackage{amsmath}
\usepackage{palatino}
\usepackage{mathpazo}
\usepackage{natbib}
\usepackage{textcomp}
\usepackage{float}
\usepackage{booktabs}
\usepackage{dcolumn}
\usepackage{booktabs}
\usepackage{multirow}
\usepackage{hyperref}
\hypersetup{colorlinks,citecolor=red}
\usepackage{amsmath}
\usepackage{xcolor}
\usepackage{orcidlink}
\usepackage[caption=false]{subfig}
\usepackage{commath}
\def\jnl@style{\it}
\def\aaref@jnl#1{{\jnl@style#1}}

\def\aaref@jnl#1{{\jnl@style#1}}

\def\aj{\aaref@jnl{AJ}}                   % Astronomical Journal
\def\apj{\aaref@jnl{ApJ}}                 % Astrophysical Journal
\def\apjl{\aaref@jnl{ApJ}}                % Astrophysical Journal, Letters
\def\apjs{\aaref@jnl{ApJS}}               % Astrophysical Journal, Supplement
\def\apss{\aaref@jnl{Ap\&SS}}             % Astrophysics and Space Science
\def\aap{\aaref@jnl{A\&A}}                % Astronomy and Astrophysics
\def\aapr{\aaref@jnl{A\&A~Rev.}}          % Astronomy and Astrophysics Reviews
\def\aaps{\aaref@jnl{A\&AS}}              % Astronomy and Astrophysics, Supplement
\def\mnras{\aaref@jnl{Mon.~Not.~Roy.~Astron.~Soc.}}             % Monthly Notices of the RAS
\def\prd{\aaref@jnl{Phys.~Rev.~D}}        % Physical Review D
\def\prc{\aaref@jnl{Phys.~Rev.~C}}  % Physical Review C
\def\prl{\aaref@jnl{Phys.~Rev.~Lett.}}    % Physical Review Letters
\def\qjras{\aaref@jnl{QJRAS}}             % Quarterly Journal of the RAS
\def\skytel{\aaref@jnl{S\&T}}             % Sky and Telescope
\def\ssr{\aaref@jnl{Space~Sci.~Rev.}}     % Space Science Reviews
\def\zap{\aaref@jnl{ZAp}}                 % Zeitschrift fuer Astrophysik
\def\nat{\aaref@jnl{Nature}}              % Nature
\def\aplett{\aaref@jnl{Astrophys.~Lett.}} % Astrophysics Letters
\def\apspr{\aaref@jnl{Astrophys.~Space~Phys.~Res.}} % Astrophysics Space Physics Research
\def\physrep{\aaref@jnl{Phys.~Rep.}}      % Physics Reports
\def\physscr{\aaref@jnl{Phys.~Scr}}       % Physica Scripta
\def\commat{\aaref@jnl{Comm.~Math.~Phys.}}              % Communications in Mathematical Physics
\def\science{\aaref@jnl{Science}}               % Science
\def\cqg{\aaref@jnl{Classical Quant.~Grav.}}            % Classical and Quantum Gravity
\def\jpcs{\aaref@jnl{JPCS}}                                     % Journal of Physics Conference Series
\def\ijmpd{\aaref@jnl{Int.~J.~Mod.~Phys.~D}}                    % International Journal of Modern Physics D
\def\grg{\aaref@jnl{Gen.~Relat.~Gravit.}}               % General Relativity and Gravitation
\def\rpp{\aaref@jnl{Rep.~Prog.~Phys.}}          % Reports on Progress in Physics
\def\npa{\aaref@jnl{Nucl.~Phys.~A}}        % Nuclear Physics A
\def\lrr{\aaref@jnl{Living Rev.~Rel.}}                   % Living reviews in relativity
\def\jcap{\aaref@jnl{J.~Cosmology Astropart.~Phys.}}    % Journal of cosmology and astroparticle physics
\def\rmp{\aaref@jnl{Rev.~Mod.~Phys.}}   %Reviews of modern physics
\def\epjc{\aaref@jnl{Eur.~Phys.~J.~C}}

\allowdisplaybreaks[1]
\begin{document}

\color{black}       %% For one column

\title{Matter Bounce Scenario in Extended Symmetric Teleparallel Gravity}

\author{A. S. Agrawal\orcidlink{0000-0003-4976-8769}}
\email{agrawalamar61@gmail.com}
\affiliation{Department of Mathematics, Birla Institute of Technology and Science-Pilani,\\ Hyderabad Campus, Hyderabad-500078, India.}
\author{B. Mishra\orcidlink{0000-0001-5527-3565}}
\email{bivu@hyderabad.bits-pilani.ac.in}
\affiliation{Department of Mathematics, Birla Institute of Technology and Science-Pilani,\\ Hyderabad Campus, Hyderabad-500078, India.}
\author{P. K. Agrawal\orcidlink{0000-0002-8976-5143}}
\email{agrawalpoonam299@gmail.com}
\affiliation{Department of Computer Sciences, Genba Sopanrao Moze, Arts Commerce and Science College, Yerwada Pune-411006, India.}

%%%%%%%%%%%%%%%%%%%%%%%%%% DATE %%%%%%%%%5%%%%%%%%%%%%%%%%%%%%%%
%%%%%%%%%%
\date{\today}

\begin{abstract}
\textbf{Abstract}:
In this paper, we have shown the matter bounce scenario of the Universe in an extended symmetric teleparallel gravity, the $f(Q)$ gravity. Motivated from the bouncing scenario and loop quantum cosmology (LQC), the form of the function $f(Q)$ has been obtained at the  backdrop of Friedmann-Lema$\hat{i}$tre-Robertson Walker (FLRW) space time. Considering the background cosmology dominated by dust fluid, the e-folding parameter has been expressed, which contains the nonmetricity term. Since the slow roll criterion in the bouncing context is not valid, we used a conformal equivalence between $f(Q)$ and scalar-tensor model to apply the bottom-up reconstruction technique in the bouncing model. The dynamics of the model has been studied through the phase space analysis, where both the stable and unstable nodes are obtained. Also, the stability analysis has been performed with the first order scalar perturbation of the Hubble parameter and matter energy density to verify the stability of the model.  

\end{abstract}

\maketitle

\section{Introduction} 

 Observational evidences suggest that the Universe had undergone an exponential expansion phase in the early Universe, known as inflation phase \cite{Brout78, Starobinsky80, Guth81}. During the inflationary phase, the Universe grew exponentially, expanded rapidly and in  a short span of time attained an immense size. The inflationary scenario has been instrumental to solve the early Universe issues like, flatness, horizon, and monopole problems. In addition, it also provides a consistent mechanism for the formation of primordial fluctuations or primordial gravitational waves. Geometrically, the expansion rate along the spatial directions can be obtained through the scale factor $a(t)$ and the evolution of Hubble parameter is based on the scale factor as, $H=\dot{a}(t)/a(t)$. So, if we look back, we could have two possibilities: (i) the scale factor attains a value zero, that leads to the big bang singularity or the space time curvature singularity, (ii) the bouncing behaviour i.e. without attain the singularity, the evolution would increase again, which is an early Universe era. Since the scale factor never zero, the space time singularity would never occur. So, according to the bouncing scenario, the Universe begins by compressing, then bounces off when it hits the minimal size of the scale factor, and begins to grow again. Hence, the bounce happens when the the value of Hubble parameter vanishes and its first derivative is positive i.e, $\dot{H}>0$.
 
 Another interesting discussion on bouncing cosmology is that it can be derived as a cosmological solution of loop quantum cosmology (LQC) \cite{Ashtekar06, Sami06, Ashtekar07, Copeland08, Corichi09, Bojowald09, Ashtekar11}. In the non-singular bouncing models, the matter bounce scenario has attracted a lot of attention. This is because the evolution of Universe even at late times comparable to a matter dominated era. Also, the matter bounce scenario generates an almost scale-invariant primordial power spectrum and leads to a matter-dominated epoch during the late phase of  expansion\cite{ Cai09, Cai13,  Quintin14, Haro15}. In this scenario, the Universe formed from an epoch in the contracting era with enormous negative time where primordial space time perturbations are generated far inside the comoving Hubble radius. The comoving Hubble radius, $r_{h} = 1/(aH)$ rises monotonically over time and eventually diverges to infinity in the far future. This has be resulted in the deceleration stage at the late expansion phase. The comoving Hubble radius in most of the bouncing models based on the modified theories of gravity grows with the cosmic time. So in far future, the decelerating age of the Universe can be experienced and it would be difficult to describe the existence of dark energy epoch.  In recent times, the bouncing scenario has been extensively studied in the curvature and torsion based modified gravity and in the scalar tensor models. \cite{ Saidov10, Barragan10, Cai11, Battefeld15, Odintsov15, Brandenberger17, Hohmann17, Ijjas18, Dombriz18, Shabani18, Matsui19, Caruana20, Mishra21, Tripathy21, Agrawal21a, Agrawal22}.\\
 
Conceptually realizing a bouncing cosmological model is not straightforward because the null energy condition has been contained in most of the phenomenological models. The null energy condition is the sum of matter pressure and energy density, which needs to be negative when the Hubble rate to grow and the bounce to happen \cite{Odintsov14} i.e. the violation of null energy condition. An exact matter bounce scenario with a single scalar field leads to an essentially scale-invariant power spectrum \cite{Odintsov20, Odintsov21}. It is noteworthy to mention here that, the matter bounce scenario is suffering from two important flaws (i) BKL (Belinski--Khalatnikov--Lifshitz) instability \cite{Belinskii70} i.e, the space time anisotropic energy density increases faster than that of the bouncing agent during the contracting phase. As a result the  background evolution became unstable; (ii)  in the perturbation evolution, large tensor to scalar ratio implying the scalar and tensor perturbations have similar amplitudes.

The extended symmetric teleparallel gravity, namely $f(Q)$ gravity is another geometrical modified theories of gravity that has been recently formulated using the non-metricity approach \cite{Jimenez18}, where $Q$ denotes the non-metricity. Several cosmological and astrophysical aspects of $f(Q)$ gravity has been studied in \cite{Lazkoz19,Bajardi20, Agrawal21b,Narawade22} . However no extensive research has been done on the matter bounce scenario in this gravitational theory. So, in this paper, we will study the matter bounce scenario motivated with the loop quantum cosmology in $f(Q)$ gravity. In Section \ref{II} we have discussed the formulation of $f(Q)$ gravity and its field equations in FLRW space time. In Section \ref{III}, the matter bounce scenario has been reconstructed in the $f(Q)$ gravity. In Section \ref{IV}, the conformal transformation has been used between $f(Q)$ and scalar-tensor model. In Section \ref{V}, the phase space analysis of the model has been performed and from the eigenvalue of the critical points, the stability of the model has been analysed. Further to show the stability behaviour of the model, the scalar perturbation analysis has been done in Section \ref{VI}. Finally, the results and discussions are given in Section \ref{VII}.    

\section{$f(Q)$ Gravity Field Equations } \label{II}
The metric tensor $g_{\mu \nu}$ is the generalisation of gravitational potential and the affine connection $\Gamma^{\mu}_{~ ~ \alpha \beta}$ describes the parallel transport and covariant derivatives. Some assumptions on the affine connection specifies the metric affine geometry \cite{Jarv18}. In differential geometry, the metric affine connection can be expressed in three independent components as \cite{Hehl95,Ortin15}, 

\begin{equation}\label{eq.1}
\Gamma^{\alpha}_{\mu \nu}=\{^{\alpha} _{~ ~\mu \nu}\}+K^{\alpha}_{~\mu \nu}+L^{\alpha}_{~\mu \nu}   
\end{equation}
where the three terms on the R.H.S. denotes the Levi-Civita Connection, Contortion and the disformation tensor respectively and can be expressed as,
\begin{eqnarray}
\{^{\alpha}_{~ ~\mu \nu}\}&\equiv&\frac{1}{2}g^{\alpha \beta}\left(\partial_{\mu}g_{\beta \nu}+\partial_{\nu}g_{\beta \mu}-\partial_{\beta}g_{\mu \nu}\right) \nonumber \\
K^{\alpha}_{~ \mu \nu}&\equiv& \frac{1}{2}T^{\alpha}_{~ \mu \nu}+T_{(\mu ~ \nu )}^{ ~ ~ \alpha}; ~~~~ T^{\alpha}_{~ \mu \nu}\equiv 2\Gamma^{\alpha}_{[\mu \nu]} \nonumber \\
L^{\alpha}_{~ \mu \nu}&\equiv&\frac{1}{2}Q^{\alpha}_{~ \mu \nu}-Q_{(\mu \nu)}^{\ \ \ \ \ \alpha} .\label{eq.2}
\end{eqnarray}

The nonmetricity conjugate is , 
\begin{equation}\label{eq.3}
P^{\alpha}_{~ ~ \mu \nu}=-\frac{1}{2}L^{\alpha}_{~  ~\mu \nu}+\frac{1}{4}\left(Q^{\alpha}-\tilde{Q}^{\alpha} \right)g_{\mu \nu}-\frac{1}{4}\delta^{\alpha}_{(\mu}Q_{\nu )}, 
\end{equation} 
where $Q_{\alpha}=g^{\mu \nu}Q_{\alpha \mu \nu}$ and $\tilde{Q}_{\alpha}=g^{\mu \nu}Q_{\mu \alpha \nu}$ with $Q_{\alpha \mu \nu}$ be the nonmetricity tensor. The  nonmetricity scalar $Q$ can be expressed as,
\begin{equation}\label{eq.4}
Q=-Q_{\alpha \mu \nu}P^{\alpha \mu \nu}
\end{equation}

The action of $f(Q)$ gravity \cite{Jimenez20} is, 
\begin{equation}\label{eq.5}
S=\int d^{4}x\sqrt{-g}\left(-\frac{1}{2}f(Q)+\mathcal{L}_{M}\right),
\end{equation}
where $g$ is the determinant of the metric $g_{\mu \nu }$; $\mathcal{L}_{M}$ be the matter Lagrangian and $Q_{\alpha \mu \nu }=\nabla_{\alpha }g_{\mu \nu }$. 

In the  geometrical framework, the flat and torsion-free connection has been considered. The connection can be parameterized with a set of functions $\xi^{\alpha}$ as
\begin{equation*}
    \Gamma^{\alpha}_{~~\mu \beta}=\frac{\partial x^{\alpha}}{\partial \xi^{\rho}}\partial_{\mu}\partial_{\beta}\xi^{\rho}
\end{equation*}
As a result, it is always feasible to make a coordinate choice that causes the connection to disappear. These coordinates are known as coincident gauge, and they are specified here as, $\mathring{\Gamma}^{\alpha}_{~\mu \nu}=0$. Thus, in the coincident gauge, $\mathring{Q}_{\alpha \mu \nu}=\partial_{\alpha}g_{\mu \nu}$, where the over ring notation refers to the coincident gauge \cite{Jimenez20}. While in the arbitrary gauge, $Q_{\alpha\mu\nu}=\partial_{\alpha}g_{\mu\nu}-2\Gamma^{\lambda}_{~~\alpha(\mu} g_{_{\nu)\lambda}}$. 

So, the field equations of $f(Q)$ gravity can be expressed as,
\begin{equation}\label{eq.6}
\frac{2}{\sqrt{-g}}\nabla_{\alpha}\left(\sqrt{-g}f_{Q}P^{\alpha}_{~ \mu \nu} \right)+\frac{1}{2}g_{\mu \nu}f+f_{Q}\left(P_{\mu \alpha \beta}Q_{\nu}^{~ \alpha \beta}-2Q_{\alpha \beta \mu }P^{\alpha \beta}_{~ ~ ~ \nu} \right)=T_{\mu \nu},
\end{equation}
where the subscript $Q$ in the function $f\equiv f(Q)$ is the partial derivative with respect to the nonmetricity scalar and the energy momentum tensor can be expressed as,
\begin{equation}\label{eq.7}
T_{\mu \nu}=-\frac{2}{\sqrt{-g}}\frac{\delta\sqrt{-g}\mathcal{L}_{m}}{\delta g^{\mu \nu}}   
\end{equation}
We shall consider here the homogeneous and isotropic FLRW space time,
\begin{eqnarray}\label{eq.8}
ds^{2}=-N^{2}(t)dt^{2}+a^{2}(t)(dx^{2}+dy^{2}+dz^{2}),
\end{eqnarray}

Where $N(t)$ and $a(t)$ respectively be the lapse function and the scale factor and the  nonmetricity scalar $Q=6\frac{H^{2}}{N^{2}}$. When the the lapse function is in standard form i.e. $N(t)=1$, then $Q=6H^2$. Since the diffeomorphism has been utilized to set the coincident gauge, therefore we can not choose any lapse function. The energy momentum tensor is that of the perfect fluid distribution and can be given as,

\begin{equation} \label{eq.9}
T_{\mu \nu}=(\rho+p)u_{\mu}u_{\nu}+pg_{\mu \nu},
\end{equation}
where the $\rho$ and $p$ respectively denotes the energy density and pressure. Now, the field equations of $f(Q)$ gravity can be obtained as, 

\begin{eqnarray} 
6f_{Q}H^{2}-\frac{1}{2}f&=&\rho  \label{eq.10} \\ 
\left(12H^{2}f_{QQ}+f_{Q}\right)\dot{H}&=&-\frac{1}{2}\left(\rho +p\right) \label{eq.11}
\end{eqnarray}
Since, the nonmetricity scalar $Q$ is directly associated with the Hubble parameter, therefore $f(Q)$ reconstruction has significant advantage over curvature based gravity, the $f(R)$ reconstruction. So, we shall reconstruct the matter bounce scenario in $f(Q)$ gravity. 

\section{$f(Q)$ Gravity in Matter Bounce Scenario} \label{III}
In the geometrical modified theories of gravity obtaining a cosmological model with bouncing scenario is not an easy task. Therefore in most of the situations the bouncing models are reconstructed based on gravitational theory. Here, we shall reconstruct such a model in the nonmetricity based gravitational theory. The focus would be mainly to reconstruct a model for which the value of Hubble squared parameter would be,  

\begin{eqnarray}\label{eq.12}
H^{2}=\frac{ \rho_{m}(\rho_{c} - \rho_{m})}{3\rho_{c}}
\end{eqnarray}

This is to mention here that the same equation can be realised from the holonomy corrected Friedmann equations in the context of LQC for a matter-dominated Universe \cite{Haro14}. The matter energy density and critical energy density are represented respectively as $\rho_{m}$ and $\rho_{c}$. Also, the critical energy density,
\begin{equation} \label{eq.13}
\rho_{c}=(c^{2}\sqrt{3})/(32\pi^{2}\gamma^{3}G_{N}l_{p}^{2}), 
\end{equation}
where,  $\gamma=0.2375$ and $l_{p}=\sqrt{\hbar G_{N}/c^{3}}$ are respectively  the Barbero-Immirzi parameter and the Planck length. Throughout this paper, we will use the Planck units, $c=\hbar=G_{N}=1$ \cite{Miranda22}. From eqn. \eqref{eq.12}, it can be inferred that when the matter energy density reaches to its critical value, $H^{2}=0$, which shows the  occurrence of a bounce. Now, in the matter bounce scenario with zero pressure, the continuity equation and the energy density can be written as,
\begin{eqnarray} \label{eq.14}
\dot{\rho}_{m}=-3H\rho_{m} \hspace{1cm} \text{and}  \hspace{1cm} \rho_{m}=\rho_{m0}a^{-3} 
\end{eqnarray}  

Motivated from the LQC, the bounce cosmology has been appealing in the sense that it can produce as a cosmological solution to the LQC theory. In some of our previous works \cite{Agrawal22}, we have used the scale factor, $a(t)\propto t^{2/3}$ for the matter dominated case. 
    
\begin{eqnarray} \label{eq.15}
\rho_{m} =\frac{{\rho_{c}}}{\left(\frac{3}{4} {\rho_{c}} t^2+1\right)}, \hspace{1.5cm} H(t)=\frac{2\rho_{c}t}{3\rho_{c}t^{2}+4},  \hspace{1.5cm} a(t)=\left({\frac{3}{4}\rho_{c} t^{2}+1}\right)^{\frac{1}{3}} 
\end{eqnarray}
Now for the above considered matter energy density the Hubble rate squared parameter becomes 
\begin{equation}\label{eq.16}
H^{2}=\frac{\rho_{c}}{3}\left(\frac{1}{a^{3}}-\frac{1}{a^{6}}\right)
\end{equation}
We have the relation between the e-folding parameter and the scale factor as, $e^{-N}=\frac{a_{0}}{a}$ \cite{Odintsov14,Nojiri09}, $a_0$ be the present value of the scale factor. Applying this in eqn. \eqref{eq.16}, we obtain,
\begin{equation}\label{eq.17}
H^{2}=\frac{\rho_{c}}{3a_{0}^{3}}\left(e^{-3N}-\frac{e^{-6N}}{a_{0}^{3}}\right)
\end{equation}
We assume following quantities,
\begin{equation}\label{eq.18}
A=\frac{\rho_{c}}{3a_{0}^{3}}, \hspace{2cm} b=\frac{1}{a_{0}^{3}}.
\end{equation}
Now, from eqn. \eqref{eq.17},  one can easily write the nonmetricity scalar in the form of e-folding parameter as,
\begin{equation}\label{eq.19}
Q=6A\left[e^{-3N}-be^{-6N}\right]
\end{equation}
On solving, the value of e-folding parameter can be obtained as,
\begin{equation}\label{eq.20}
N=-\frac{1}{3}Log\left(\frac{3A+\sqrt{9A^{2}-6AbQ}}{6Ab}\right)
\end{equation}
In addition, we assume that the energy density \eqref{eq.10} is of form,
\begin{equation}
\rho=\sum_{i}\rho_{i0}a_{0}^{-3(1+\omega_{i})}e^{-3N(1+\omega_{i})}    
\end{equation}\label{eq.21}
By setting $S_{i}=\rho_{i0}a_{0}^{-3(1+\omega_{i})}$, the energy density becomes
\begin{equation}\label{eq.22}
\rho=\sum_{i}S_{i}\left(\frac{3A+\sqrt{9A^{2}-6AbQ}}{6Ab}\right)^{(1+\omega_{i})}
\end{equation}
Substituting eqn. \eqref{eq.22} in eqn. \eqref{eq.10}, we get  
\begin{equation}\label{eq.23}
Qf_{Q}-\frac{1}{2}f-\sum_{i}S_{i}\left(\frac{3A+\sqrt{9A^{2}-6AbQ}}{6Ab}\right)^{(1+\omega_{i})}=0
\end{equation}

\begin{figure}[H]
\centering
\minipage{0.40\textwidth}
\includegraphics[width=\textwidth]{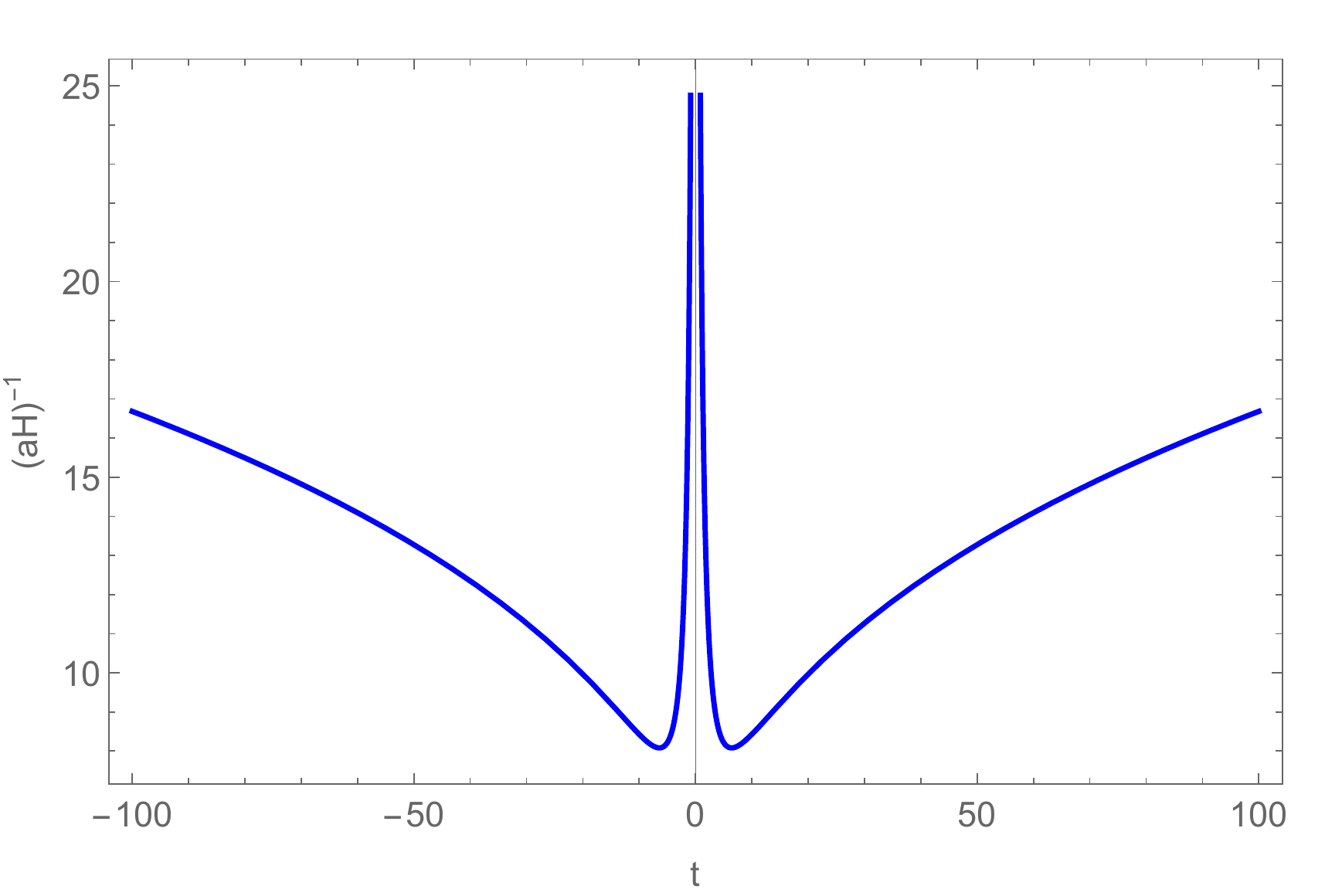}
\endminipage\hfill
\caption{Evolutionary behaviour of Hubble radius in cosmic time.}\label{FIG.1}
\end{figure}

Since we consider the Universe is filled with dust fluid only; the pressure term becomes zero which implies that the equation of state parameter vanishes. From eqn. \eqref{eq.18} one can easily find the value of matter-energy density at $\omega_{i}=0$ as, 
\begin{equation}\label{eq.24}
Qf_{Q}-\frac{1}{2}f-\left(\frac{\rho_{c}+\sqrt{\rho_{c}(\rho_{c}-2Q)}}{2}\right)=0
\end{equation}
On solving, we get
\begin{equation}\label{eq.25}
f(Q)=-\sqrt{\rho_{c} (\rho_{c}-2 Q)}-\sqrt{2\rho_{c} Q} \arcsin{\left(\frac{\sqrt{2} \sqrt{Q}}{\sqrt{\rho_{c}}}\right)}-\rho_{c},
\end{equation}
The above form of $f(Q)$ produces the matter bounce evolution of the Universe. In bouncing cosmology, the scale factor contracts in the pre-bounce epoch, then increases after reaching the minimum value at $t=0$, and in the post-bounce epoch, its evolution shows symmetric behaviour to that of pre-bounce. The Hubble parameter traverses from $H<0$ to $H>0$ and crosses $H=0$ at $t=0$. The late-time acceleration of the Universe epoch is ensured by the diminishing trend of the cosmic Hubble radius as shown in FIG. \ref{FIG.1}. Furthermore, the primordial perturbation modes generate during the deep contracting era far away from the bounce, when all perturbation modes lie within the horizon, due to the growth of the Hubble radius. However, in this case, the Hubble parameter diverges as we move away from the bounce epoch, which shows the decelerating behavior of the Universe at a late time.

\begin{figure}[H]
\centering
\minipage{0.40\textwidth}
\includegraphics[width=\textwidth]{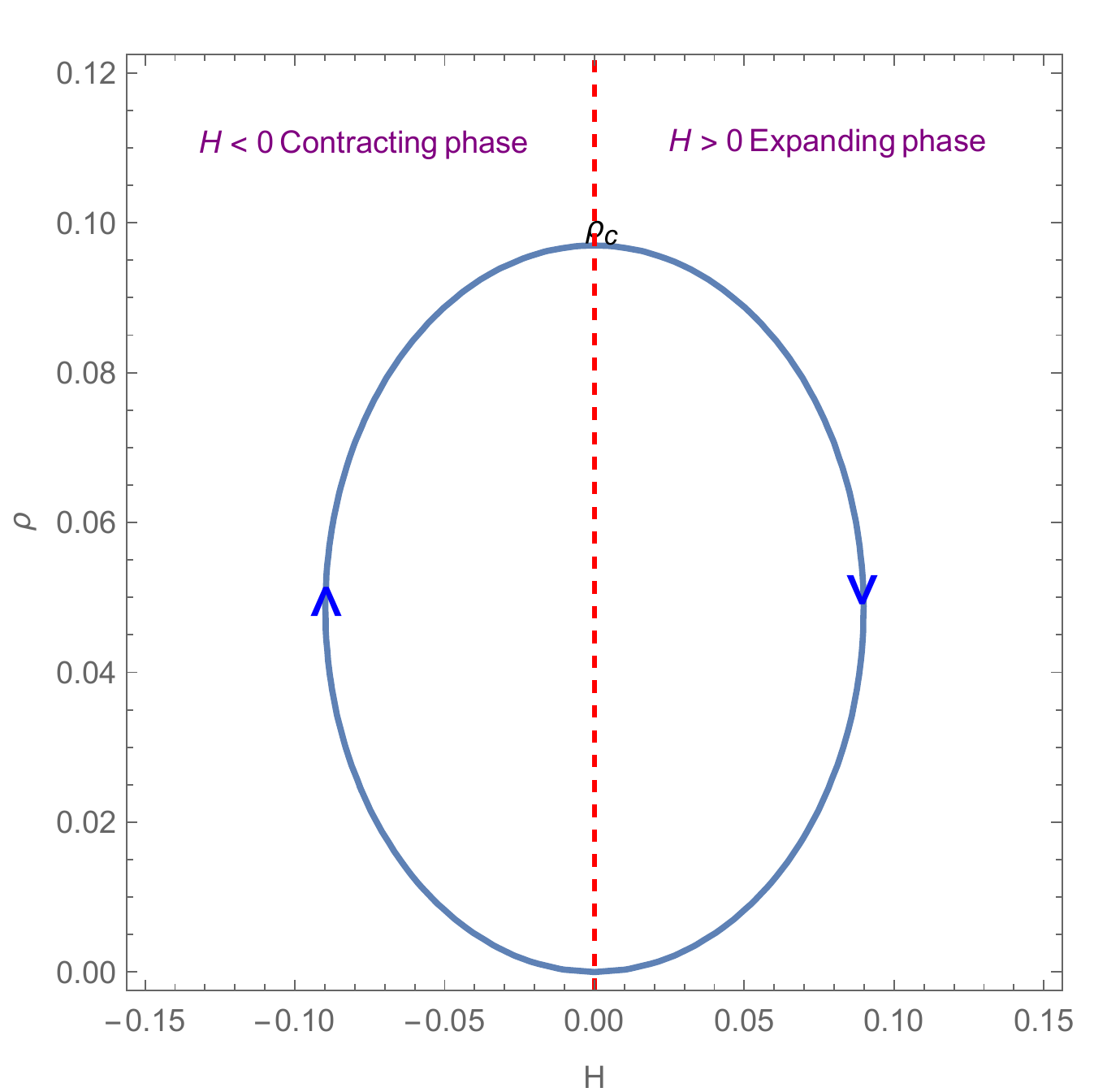}
\endminipage\hfill
\minipage{0.40\textwidth}
\includegraphics[width=\textwidth]{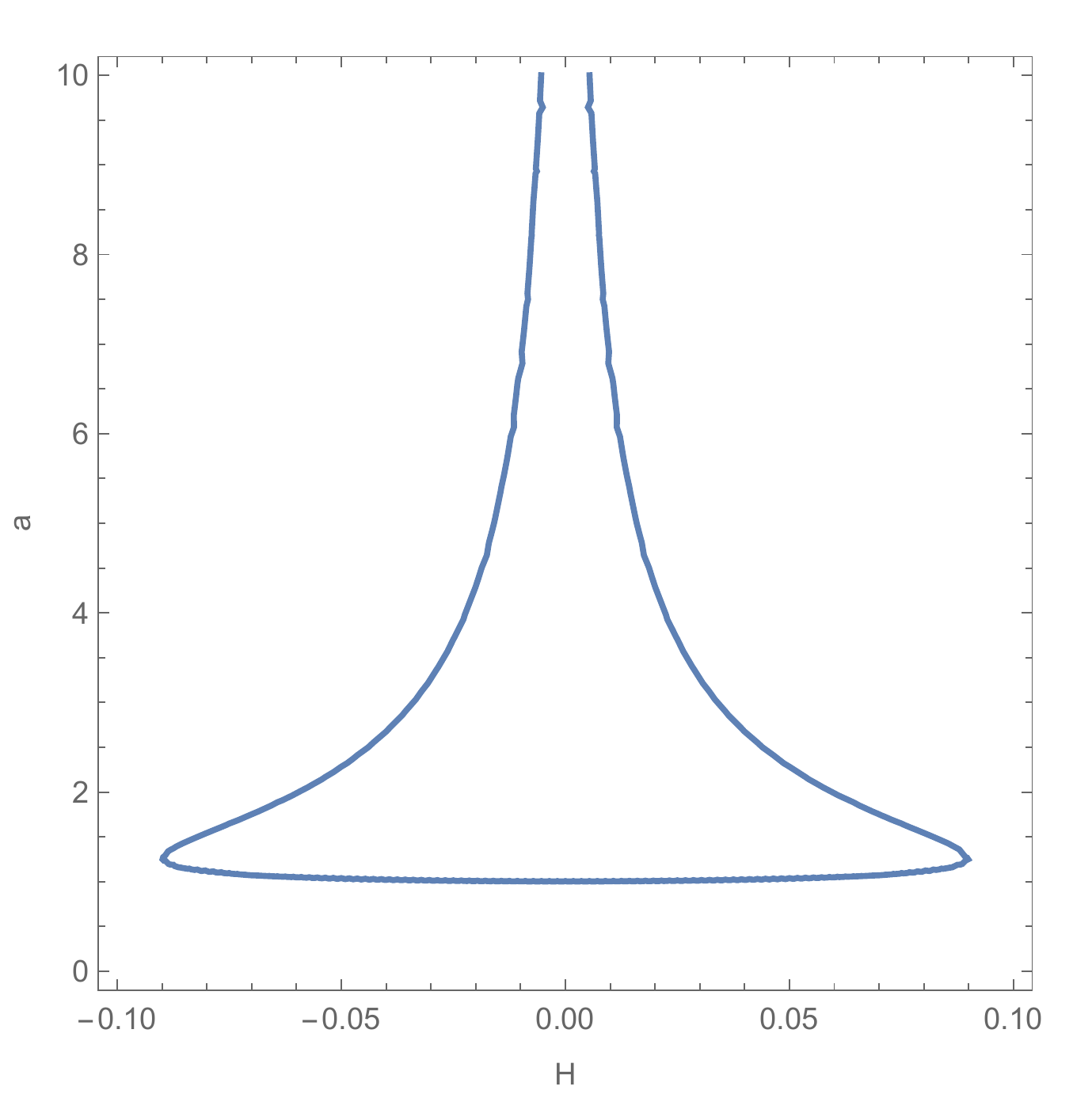}
\endminipage
\caption{Evolutionary behaviour of energy density (left panel) and scale factor (right panel) in Hubble parameter.}\label{FIG.2}
\end{figure}

The evolutionary behaviour of energy density and scale factor with respect to Hubble parameter has been shown in FIG.  \ref{FIG.2}. The elliptic curve in the right half of  FIG.  \ref{FIG.2} (left panel) shows the expanding universe in which the Hubble parameter remains positive. Whereas, left side shows the contracting phase of the universe. It has also been seen that, the Hubble parameter attains zero value (red dotted line) at the minimum and maximum value of the energy density, respectively at $0$ and $\rho_{c}$ \cite{Amoros13}. Referring eqn. \eqref{eq.14}, we can observe that the Universe traverses clockwise along the ellipse from contracting to expanding phase. This behaviour is because of the fact that the energy density increases in the contracting phase and in the expanding phase, it decreases. FIG. \ref{FIG.2} (right panel) shows the behaviour of scale factor with respect to the Hubble parameter. When the scale factor reaches its minimum value the Hubble parameter vanishes. The value of the scale decreases during the  contracting phase of the universe, then bounces back to its minimum before increasing during the expansion phase of the universe.
%%%%%%%%%%%%%%%%%%%%%%%%%%%%%%%%%%%%
\section{Conformal Transformation}\label{IV}
Let us take the scalar field $\varphi$ as independent variable and consider the action functional \cite{Jarv18},
\begin{equation}\label{eq.26}
S=\frac{1}{2}\int d^{4}x\sqrt{-g}\left[Q-\mathcal{B}(\varphi)g^{\alpha \beta}\partial_{\alpha}\varphi \partial_{\beta}\varphi -2\mathcal{V}(\varphi)+2\lambda_{\mu}^{~\beta \alpha \gamma}R^{\mu}_{~\beta \alpha \gamma}+2\lambda_{\mu}^{~\alpha \beta}T^{\mu}_{~\alpha \beta}\right]
\end{equation} 
$\mathcal{B}(\varphi)=\mathcal{V}(\varphi)=0$ is the conditions under which the theory simplifies to symmetric teleparallel equivalent to general relativity (STEGR). The antisymmetry of the linked geometrical objects is presumptively respected by the Lagrange multipliers. Therefore, vanishing curvature $R^{\mu}_{~\beta \alpha \gamma}=0$ and torsion $T^{\mu}_{~\alpha \beta}=0$ are imposed, as is expected in the symmetric teleparallel framework, by the equations $\lambda_{\mu}^{~\beta \alpha \gamma}=\lambda_{\mu}^{~\beta [\alpha \gamma]}$ and $\lambda_{\mu}^{~\alpha \beta}=\lambda_{\mu}^{~[\alpha \beta]}$.   
   
Contrary to the scalar-curvature case the action does not preserve its form under the local conformal rescaling of the metric. 
\begin{equation}\label{eq.27}
\tilde{g}_{\mu \nu}=e^{-\sqrt{2/3}\Omega(\varphi)}g_{\mu \nu}
\end{equation} 
%The non-metricity scalar transform under the above conformal transformation as follows:
% \begin{equation}\label{eq.28}
% \tilde{Q}=e^{\sqrt{2/3}\Omega(\varphi)}\left[Q+\left(\frac{d\Omega}{d\varphi}\right)^{2}g^{\alpha \beta} \partial_{\alpha}\varphi \partial_{\beta}\varphi +\frac{d\Omega}{d\varphi}(Q^{\alpha}-\tilde{Q}^{\alpha})\partial_{\alpha}\varphi \right]
% \end{equation}
The non-metricity scalar in the $f(Q)$ and scalar-tensor frames, respectively, are $Q$ and $\tilde{Q}$.
%%%%%%%%%%%%%%%
% \begin{equation*}
% Q=e^{-\sqrt{2/3}\Omega(\varphi)}\tilde{Q}-\left(\frac{d\Omega}{d\varphi}\right)^{2}g^{\alpha \beta} \partial_{\alpha}\varphi \partial_{\beta}\varphi -\frac{d\Omega}{d\varphi}(Q^{\alpha}-\tilde{Q}^{\alpha})\partial_{\alpha}\varphi 
% \end{equation*}
%%%%%%%%%%%%%%%%%
%%%%%%%%%%%%%%%
% \begin{equation*}
% Q=e^{-\sqrt{2/3}\Omega(\varphi)}\left[\tilde{Q}-e^{\sqrt{2/3}\Omega(\varphi)}\left(\frac{d\Omega}{d\varphi}\right)^{2}g^{\alpha \beta} \partial_{\alpha}\varphi \partial_{\beta}\varphi -e^{\sqrt{2/3}\Omega(\varphi)}\frac{d\Omega}{d\varphi}(Q^{\alpha}-\tilde{Q}^{\alpha})\partial_{\alpha}\varphi \right]
% \end{equation*}
%%%%%%%%%%%%%%%%%
\begin{equation}\label{eq.28}
Q=e^{-\sqrt{2/3}\Omega(\varphi)}\left[\tilde{Q}-\left(\frac{d\Omega}{d\varphi}\right)^{2}\tilde{g}^{\alpha \beta} \partial_{\alpha}\varphi \partial_{\beta}\varphi -e^{\sqrt{2/3}\Omega(\varphi)}\frac{d\Omega}{d\varphi}(Q^{\alpha}-\tilde{Q}^{\alpha})\partial_{\alpha}\varphi \right]
\end{equation}
%%%%%%%%%%%%%%%%%
\begin{equation}\label{eq.29}
S=\frac{1}{2}\int d^{4}x\sqrt{-\tilde{g}}\left[\tilde{Q}-\mathcal{B}\tilde{g}^{\alpha \beta}\partial_{\alpha}\varphi \partial_{\beta}\varphi -2\mathcal{V}(\varphi)\right]
\end{equation}
%%%%%%%%%%%%%%%%%
Due to the fact that $Q^{\alpha}=Q^{\alpha \mu}_{~~~\mu}$ and $\tilde{Q}^{\alpha}\equiv Q_{\mu}^{~\mu \alpha}$, respectively. The component $(Q^{\alpha}-\tilde{Q}^{\alpha})\partial_{\alpha}\Omega$ does not exist in the initial action, but the additional piece proportional to $\left(\frac{d\Omega}{d\varphi}\right)^{2}\tilde{g}^{\alpha \beta} \partial_{\alpha}\varphi \partial_{\beta}\varphi$ can be absorbed into the redefinition of the kinetic term, $\mathcal{B}(\varphi)=\left(\frac{d\Omega}{d\varphi}\right)^{2}$ of the scalar field. 

If the $f(Q)$ model space-time is characterized by a FRW metric with $\eta$ be the conformal time and $a(\eta)$ is the scale factor i.e., 

\begin{equation}\label{eq.30}
ds^{2}=a^{2}(\eta)[-d\eta^{2}+\delta_{\mu \nu}dx^{\mu}dx^{\nu}]
\end{equation} 
 then the associated scalar-tensor model's metric changes to
\begin{eqnarray}
d\tilde{s}^{2}&=& e^{-\sqrt{2/3}\Omega(\varphi)}a^{2}(\eta)[-d\eta^{2}+\delta_{\mu \nu}dx^{\mu}dx^{\nu}] \nonumber \\
&=& a_{s}^{2}(\eta)[-d\eta^{2}+\delta_{\mu \nu}dx^{\mu}dx^{\nu}]\label{eq.31}
\end{eqnarray}
In the scalar-tensor model, $a_{s}(\eta)=e^{-\sqrt{1/6}\Omega(\varphi)}a(\eta)$ is the scale factor. The conformal time is unchanged in both frames, but the cosmic time undergoes a transformation via the formula $dt_{s}=e^{-\sqrt{1/6}\Omega(\varphi)}dt$. Notation will be used in this paper : the cosmic time, scale factor in the $f(Q)$ frame and scalar-tensor frame are represented by the symbols ($t,a(t)$) and ($t_{s}, a_{s}(t_{s})$) respectively. In terms of the Hubble parameter, $H$ stands for the $f(Q)$ frame and $H_{s}$ for the scalar-tensor. 
 
One can get the field equations in the scalar tensor frame,
\begin{eqnarray}
3H_{s}^{2}&=&\frac{1}{2}\mathcal{B}\left(\frac{d\varphi}{dt_{s}}\right)^{2}+\mathcal{V} \label{eq.32}\\
2\frac{dH_{s}}{dt_{s}}+3H_{s}^{2}&=&-\frac{1}{2}\mathcal{B}\left(\frac{d\varphi}{dt_{s}}\right)^{2}+\mathcal{V} \label{eq.33}
\end{eqnarray}

Using the scalar field equation we get

\begin{equation}\label{eq.34}
\mathcal{B}\frac{d^{2}\varphi}{dt_{s}^{2}}+3H_{s}\mathcal{B}\frac{d\varphi}{dt_{s}}+\frac{1}{2}\frac{d\mathcal{B}}{dt_{s}}\frac{d\varphi}{dt_{s}}+\frac{d\mathcal{V}}{d\varphi}=0
\end{equation}

where $\frac{d}{dt_{s}}=\frac{1}{a_{s}(\eta)}\frac{d}{d\eta}$.

From the preceding equations, one can readily obtain $2\frac{dH_{s}}{dt_{s}}=-\mathcal{B}\left(\frac{d\varphi}{dt_{s}}\right)^{2}$. Because we are dealing with an inflationary scenario, the slow roll conditions still apply in the scalar tensor model. The slow roll condition is created by inputting a few slow roll characteristics that are believed to be less than unity during an inflationary period.
\begin{equation}\label{eq.35}
\epsilon_{1}=-\frac{1}{H_{s}^{2}}\frac{dH_{s}}{dt_{s}}, \hspace{1cm} \epsilon_{2}=\frac{1}{H_{s}}\frac{d^{2}\varphi/dt_{s}^{2}}{d\varphi/dt_{s}}, \hspace{1cm} \epsilon_{4}=\frac{1}{2\mathcal{B}H_{s}}\frac{d\mathcal{B}}{dt_{s}}
\end{equation}  

There is another slow roll parameter defined as $\epsilon_{3}=\frac{1}{2H_{s}\mathcal{G}_{\tilde{Q}}}\frac{d\mathcal{G}_{\tilde{Q}}}{dt_{s}}, ~\mathcal{G}_{\tilde{Q}}=\partial \mathcal{G}/\partial \tilde{Q}$ in more general actions like $S=\frac{1}{2}\int d^{4}x\sqrt{-\tilde{g}}\left[\mathcal{G}(\tilde{Q},\varphi)-\mathcal{B}\tilde{g}^{\alpha \beta}\partial_{\alpha}\varphi \partial_{\beta}\varphi -2\mathcal{V}(\varphi)\right]$ (where $\mathcal{G}(\tilde{Q},\varphi)$ is any analytic function of $\tilde{Q}$ and $\varphi$), but in the current situation, i.e., for action eq. \eqref{eq.29} $\mathcal{G}(\tilde{Q},\varphi)=\tilde{Q}$, and thus the slow roll parameter $\epsilon_{3}$ vanishes \cite{Nojiri17, Hwang05}. With the condition $\epsilon_{i}\ll 1$, the spectral index for curvature perturbation and the tensor to scalar ratio of the ST model are given by

\begin{eqnarray}\label{eq.36}
n_{s}&=&1-4\epsilon_{1}-2\epsilon_{2}-2\epsilon_{4} \nonumber \\
r&=&\frac{8\mathcal{B}}{H_{s}^{2}}\left(\frac{d\varphi}{dt_{s}}\right)^{2}
\end{eqnarray}
The gravitational equation $\frac{2dH_{s}}{dt_{s}}=-\mathcal{B}\left(\frac{d\varphi}{dt_{s}}\right)^{2}$ provides the following simplified form of the tensor to scalar ratio:
\begin{equation}\label{eq.37}
r=-\frac{16}{H_{s}^{2}}\frac{dH_{s}}{dt_{s}}=16\epsilon_{1}
\end{equation}
Furthermore, due to the slow roll conditions, the equations of motion can be approximated as follows.
\begin{equation}\label{eq.38}
3H_{s}^{2}\simeq \mathcal{V}(\varphi)
\end{equation}
and 
\begin{equation}\label{eq.39}
\frac{1}{2}\frac{d\mathcal{B}}{d\varphi}\left(\frac{d\varphi}{dt_{s}}\right)^{2}+3H_{s}\mathcal{B}\frac{d\varphi}{dt_{s}}+\frac{d\mathcal{V}}{d\varphi}=0
\end{equation}
   
Having set the stage, let us consider an ansatz of tensor-to-scalar ratio in terms of the e-folding number as,
\begin{equation}\label{eq.40}
r(N_{s})=16e^{\beta(N_{s}-N_{f})}
\end{equation} 
 
where $\beta$ is a model parameter with no dimensions and $N_{s}$ is the e-folding parameter in ST frame. It should be noted that the e-folding number can be defined as either $N_{s}=\int_{t_{h}}^{t_{s}}{H_{s}dt_{s}}$ or $N_{s}=\int_{t_{s}}^{t_{\text{end}}}{H_{s}dt_{s}}$, where $t_{h}$ and $t_{\text{end}}$ are the onset and end points of the inflation, respectively. In the former situation, $dN_{s}/dt_{s}>0$, the e-folding parameter increases monotonically with the cosmic time $t_{s}$, whereas in the later instance, $dN_{s}/dt_{s}<0$, the e-folding parameter drops monotonically with the cosmic time $t_{s}$. The most crucial component is to see if the $r(N_{s})$ decision results in observable conformity with the Planck restrictions. Using the relation $\frac{d}{dt_{s}}=H_{s}\frac{d}{dN_{s}}$, we may compare equations \eqref{eq.37} and \eqref{eq.40}. 
\begin{equation}\label{eq.41}
\frac{1}{H_{s}}\frac{dH_{s}}{dN_{s}}=-e^{\beta(N_{s}-N_{f})}  
\end{equation} 
The Hubble parameter in the form of e-folding parameter can be written as 
\begin{equation}\label{eq.42}
H_{s}(N_{s})=H_{s0}~Exp\left(-\frac{1}{\beta}e^{\beta(N_{s}-N_{f})}\right)
\end{equation}
It is noteworthy to mention here that the ansatz that is considered in equation \eqref{eq.40} allows an inflationary scenario of the universe having an exit at $N_{s}=N_{f}$ ie., at $t_{s}=t_{\text{end}}$. On the other hand, near the beginning of the inflation the Hubble parameter follows a quasi de-sitter evolution \cite{Odintsov20}.

Using the relation $\frac{dN_{s}}{dt_{s}}=H_{s}(N_{s})$ the conformal time can be defined as follow,
\begin{equation}\label{eq.43}
\eta(N_{s})=-\frac{Exp\left(\frac{1}{\beta}e^{-\beta N_{f}}\right)}{H_{s0}(1-e^{-\beta N_{f}})}e^{-(1-e^{-\beta N_{f}})N_{s}}
\end{equation} 
The next step is to find the conformal factor $\Omega(\varphi)$ in such a way that the conformally transformed $f(Q)$ frame scale factor results in a non-singular bounce after the inflationary scenario in the scalar tensor frame has been verified. We select 
\begin{equation}\label{eq.44}
\Omega(\varphi(N_{s}))=\sqrt{6}~ln\left[e^{-N_{s}}\left(\frac{3}{4}\rho_{c}\eta^{2}(N_{s})+1\right)^{\frac{1}{3}}\right]
\end{equation}  
It is simple to see that the conformally connected $f(Q)$ frame scale factor exhibits the following behaviour because of the aforesaid form of $\Omega(\varphi)$.
\begin{equation}\label{eq.45}
a(\eta)=\left(\frac{3}{4}\rho_{c}\eta^{2}+1\right)^{\frac{1}{3}}
\end{equation}
It is simple to demonstrate that the scale factor indicated above causes a non-singular bounce at $\eta=0$. Additionally, close to $\eta=0$, the $f(Q)$ frame scale factor can be approximated as $a(\eta)=1+\frac{1}{4}\rho_{c}\eta^{2}$, and as a result, the conformal time is connected to the $f(Q)$ cosmic time by $t=\int a(\eta)d\eta =\eta+\frac{\rho_{c}\eta^{3}}{12}\approx \eta$. Because of this, the scale factor in terms of cosmic time turns out to be $a(t)=\left(\frac{3}{4}\rho_{c}t^{2}+1\right)^{\frac{1}{3}}$ in $f(Q)$ frame.
 
Using the relation $\mathcal{B}(\varphi)=\left(d\Omega/d\varphi\right)^{2}$ the spectral index can be defined as follows
\begin{equation}\label{eq.46}
n_{s}=1+\frac{4}{H_{s}^{2}}\frac{dH_{s}}{dt_{s}}+\frac{2\left(3\frac{dH_{s}}{dt_{s}}\mathcal{B}+3H_{s}\frac{d\mathcal{B}}{dt_{s}}+\frac{1}{2}\frac{d^{2}\mathcal{B}}{dt_{s}^{2}}\right)}{H_{s}\left(3H_{s}\mathcal{B}+\frac{1}{2}\frac{d\mathcal{B}}{dt_{s}}\right)}-\frac{1}{\mathcal{B}H_{s}}\frac{d\mathcal{B}}{dt_{s}}
\end{equation} 
We will determine the scalar spectral index in terms of e-folding number and for this reason we needs the following identities;
\begin{equation}\label{eq.47}
\frac{d}{dt_{s}}=H_{s}\frac{d}{dN_{s}}, \hspace{1cm} \frac{d^{2}}{dt_{s}^{2}}=H_{s}^{2}\frac{d^{2}}{dN_{s}^{2}}+H_{s}\frac{dH_{s}}{dN_{s}}\frac{d}{dN_{s}}
\end{equation}
 To determine the value of spectral index the relation $\mathcal{B}(\varphi)=\left(d\Omega/d\varphi\right)^{2}$ can be used which provide the right hand side of the eq. (\ref{eq.46}) as
 \begin{align}
n_{s}=1-2e^{\beta (N_{s}-N_{f})}-2\frac{d^{2}\Omega}{dN_{s}^{2}}\left(\frac{d\Omega}{dN_{s}}\right)^{-1}+\frac{2}{\left(\frac{d^{2}\Omega}{dN_{s}^{2}}+(3-e^{\beta(N_{s}-N_{f})\frac{d\Omega}{dN_{s}}})\right)}\nonumber \\
\times \left[-3\frac{d\Omega}{dN_{s}}e^{\beta(N_{s}-N_{f})}+\left(\frac{d^{2}\Omega}{dN_{s}^{2}}-\frac{d\Omega}{dN_{s}}e^{\beta(N_{s}-N_{f})}\right)\left(6-3e^{\beta(N_{s}-N_{f})}+\frac{d^{2}\Omega}{dN_{s}^{2}}\right)
\right. \nonumber \\ \left. 
+\left(\frac{d^{3}\Omega}{dN_{s}^{3}}-\frac{d^{2}\Omega}{dN_{s}^{2}}e^{\beta(N_{s}-N_{f})}-\beta\frac{d\Omega}{dN_{s}}e^{\beta(N_{s}-N_{f})}\right)\right]\label{eq.48}
 \end{align}
The integral can be performed for the limit $N_{s}\rightarrow{0}$ in equation (\ref{eq.43}) i.e., near the horizon crossing time, which is sufficient in the current context because observable quantities such as spectral index and tensor to scalar ratio are eventually determined at the horizon crossing instance. As a result, the conformal factor in terms of the e-folding number takes the following form:
\begin{equation}\label{eq.49}
\Omega({N_{s}})=\sqrt{6}\left[-N_{s}+ln\left(\frac{3}{4}\rho_{c}\left(\frac{e^{\frac{1}{\beta}e^{-\beta N_{f}}-(1-e^{-\beta N_{f}})N_{s}}}{H_{s0}(1-e^{-\beta N_{f})}}\right)^{2}+1\right)^{\frac{1}{3}}\right]   
\end{equation}
The dimensionless parameter $\beta$ determines the tensor-to-scalar ratio in eq. (\ref{eq.40}), and $N_{f}-N(t_{h})=N_{T}$, where $N_{T}$ is the total e-folding of the inflationary epoch and $t_{h}$ is the horizon crossing instance. For $\beta>0.092$, the tensor-to-scalar ratio is inside the Planck restrictions for $N_{T}=60$. So, for $\beta=0.1$ and $N_{T}=60$, the spectral index derived in eq. (\ref{eq.48}) is consistent with Planck results. Now from eqs. (\ref{eq.40}), (\ref{eq.48}) and (\ref{eq.49}) values for scalar spectral index and the tensor to scalar ratio in the scalar tensor frame are $n_{s}=0.9649\pm 0.0042$ and $r<0.064$ respectively from the Planck 2018 constraints \cite{Akrami20}.
%%%%%%%%%%%%%%%%%%%%%%%%%%%%%%%%%%%%
\section{Phase Space Analysis} \label{V}
The phase space analysis is the study where all possible states of the system have been represented and each possible state has a unique point. This can also be described as the combination of all possible values of position space and momentum space. We will perform the phase space analysis of the system that we have obtained in the form of $f(Q)$ as in eqn.  \eqref{eq.25}. We consider here a general form of $f(Q)$ as $Q+\psi(Q)$ \cite{Narawade22, Khyllep21} and accordingly eqns. \eqref{eq.10}, \eqref{eq.11} take the form,
\begin{eqnarray} 
3H^{2}&=&\rho+\frac{\psi}{2}-Q\psi_{Q} \label{eq.50} \\  
2\dot{H}+3H^{2}&=&-p-2\dot{H}(2Q\psi_{QQ}+\psi_{Q})+\left(\frac{\psi}{2}-Q\psi_{Q}\right)    \label{eq.51}
\end{eqnarray}   
When the Universe comprises of matter and radiation fluids, one can obtain the following relation 
\begin{equation} \label{eq.52}
\rho=\rho_{m}+\rho_{r}, \hspace{2cm}  p=\frac{\rho_{r}}{3}  ,  
\end{equation}
with the matter and radiation energy density represented respectively as  $\rho_{m}$ and $\rho_{r}$. Hence, we can have the relation
\begin{eqnarray} 
3H^{2}&=&\rho+\rho_{de}  \label{eq.53} \\ 
2\dot{H}+3H^{2}&=&-p-p_{de},    \label{eq.54}
\end{eqnarray}

Comparing eqns. \eqref{eq.50} with \eqref{eq.53} and \eqref{eq.51} with \eqref{eq.54}, the dark energy density and dark energy pressure contributions caused by the geometry can be separated as,
\begin{eqnarray}
\rho_{de}&=&\frac{\psi}{2}-Q\psi_{Q} \label{eq.55} \\    
p_{de}+\rho_{de}&=&2\dot{H}(2Q\psi_{QQ}+\psi_{Q}) \label{eq.56}   
\end{eqnarray}

The density parameters for the matter dominated, radiation dominated and dark energy phase are respectively denoted as, $ \Omega_{m}=\frac{\rho_{m}}{3H^{2}}$, $\Omega_{r}=\frac{\rho_{r}}{3H^{2}}$ and $\Omega_{de}=\frac{\rho_{de}}{3H^{2}}$ with $\Omega_{m}+\Omega_{r}+\Omega_{de}=1$. Hence the effective equation of state parameter takes the form.
\begin{eqnarray}
\omega_{eff}&=&-1+\frac{\Omega_{m}+\frac{4}{3}\Omega_{r}}{2Q\psi_{QQ}+\psi_{Q}+1}   \label{eq.57} 
\end{eqnarray}

So, to analyze the dynamics of the model, we consider the dimensionless variables, $x=\frac{\psi-2Q\psi_{Q}}{6H^{2}}$ and $ y=\frac{\rho_{r}}{3H^{2}}$, which has been transformed into an autonomous dynamical system. Further, if prime denotes the differentiation for the number of e-folds of the Universe $N=ln a$, then the equations of the model can be computed using the chain rule as,
\begin{equation}
\phi'=\frac{d\phi}{dN}=\frac{d\phi}{dt}\frac{dt}{da}\frac{da}{dN}=\frac{\dot{\phi}}{H}     \label{eq.58}
\end{equation}
 
So, the autonomous dynamical system can be given as, 
\begin{eqnarray}
x'&=&-2\frac{\dot{H}}{H^{2}}(\psi_{Q}+2Q\psi_{QQ}+x) \label{eq.59}\\
y'&=&-2y\left(2+\frac{\dot{H}}{H^{2}}\right), \label{eq.60}
\end{eqnarray}
and with an algebraic manipulation, we can obtain the relation, $
 \frac{\dot{H}}{H^{2}}=-\frac{1}{2}\left(\frac{3-3x+y}{2Q\psi_{QQ}+\psi_{Q}+1}\right)$. If we compare the $f(Q)$ as obtained in eqn. \eqref{eq.25}, then $\psi(Q)$ can be represented as,  
\begin{equation}\label{eq.61}
\psi(Q)=-\sqrt{\rho_{c} (\rho_{c}-2 Q)}-\sqrt{2\rho_{c} Q} \arcsin \left(\frac{\sqrt{2} \sqrt{Q}}{\sqrt{\rho_{c}}}\right)-\rho_{c}-Q    
\end{equation}
and 
\begin{equation} \label{eq.62}
2Q\psi_{QQ}+\psi_{Q}=-\frac{\rho_{c}}{\sqrt{\rho_{c} (\rho_{c}-2 Q)}}-1   
\end{equation}
Now the dimensionless variables can be represented as,
\begin{eqnarray}
x'&=&x\left(3(x-1)-y\right) \label{eq.63} \\
y'&=&-\frac{y (x (-3 x+y+4)+y-1)}{x-1} \label{eq.64}
\end{eqnarray}

\begin{figure}[H]
\centering
\minipage{0.40\textwidth}
\includegraphics[width=\textwidth]{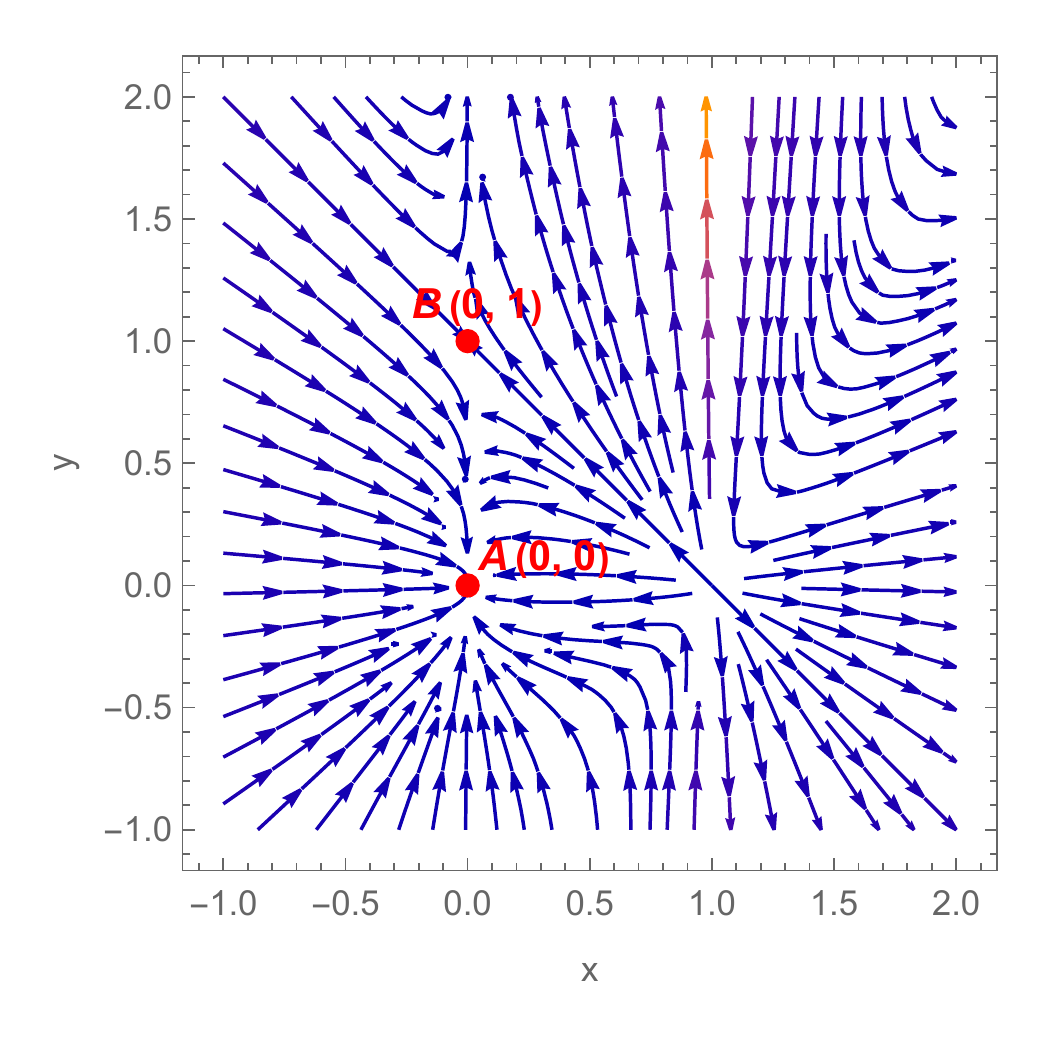}
\endminipage\hfill
\caption{Phase-space trajectories on the $x$-$y$ plane for $f(Q)$ gravity.}\label{FIG.3}
\end{figure}

The critical points of the above system of equations are $(0,0)$, $(0,1)$ and the stability can be checked from their corresponding eigenvalues. We obtained the eigenvalues $\{-3,-1\}$ corresponding to the critical point $(0,0)$ whereas $\{-4,1\}$ for the critical point $(0,1)$. Since, both the eigenvalues at the critical point $(0,0)$ are negative it implies the stable node.  On the other hand at $(0,1)$, the eigenvalues contains both positive and negative real part hence it implies unstable at $(0,1)$.  The effective equation of state [Eqn. \eqref{eq.57}] and the deceleration parameter can be obtained in terms of the dynamical variables respectively as,
\begin{eqnarray}
\omega_{eff}&=&-1+\frac{(x+1) (3 x-y-3)}{3 (x-1)} \label{eq.65} \\
q&=&-1+\frac{(x+1) (3 x-y-3)}{2 (x-1)}    \label{eq.66}  
\end{eqnarray}
The details of the critical points and its behaviour are given in the following phase portrait (FIG. \ref{FIG.3}) and the corresponding cosmology in TABLE I.

\begin{table}[ht] \label{TABLE I}
\caption{Critical points for the dynamical system of the considered model}
\begin{center}
\begin{tabular}{ |c|c|c|c|c|c|c|c|c } 
 \hline
 Point($x,y$)  & $\Omega_{m}$ & $\Omega_{r}$ &  $\Omega_{de}$ & $\omega_{eff}$ & Deceleration $(q)$ & Eigenvalues& Stability   \\ \hline
$A(0,0)$ & 1 & 0 & 0 &  0 &  1/2 & \{-3,-1\} &Stable Node \\ \hline
$B(0,1)$ & 0 & 1 & 0 & 1/3 &  1 & \{-4,1\}&Unstable \\ \hline
\end{tabular}
\end{center}
\end{table}

The cosmological properties of the model can be extracted  in the dynamical system approach without obtaining the exact solution to the evolution equations. In addition, the information on the gravitational theory and the cosmic evolution can also be analysed. The cosmic dynamics of the model can be analysed by obtaining the critical points after solving the systems that comprises of $x$ and $y$. In TABLE- I,  it is shown that the system comprised of two critical points, with one among them shows the stable behaviour. The point $A(0,0)$ corresponds to a matter-dominated Universe, and $\omega_{eff}=0$ indicates the same that the Universe is in a matter-dominated phase, as shown in the phase space portrait also. The position $B(0,1)$ corresponds to a radiation-dominated phase; the $\omega_{eff}=1/3$ indicates the radiation-dominated era. Moreover, the deceleration parameter is positive at both points, showing deceleration behavior. It is noteworthy to mention here that in most of the research on bouncing cosmology, it has been reported that the matter bounce scenario fails to explain the late time dark energy era and in this case also no critical points are obtained which indicates the dark energy era.
%%%%%%%%%
\section{Stability Analysis with Scalar Perturbation} \label{VI}
We shall undertake the scalar perturbation analysis to discuss the stability behaviour of the reconstructed bouncing scenario model in $f(Q)$ gravity. We shall adhere to the linear homogeneous and isotropic perturbation and will describe the perturbation of energy density and Hubble parameter \cite{Wu12,Golovnev18,Duchaniya22}. The first order perturbation in the FLRW background with the perturbation geometry functions $\delta(t)$ and matter functions $\delta_m(t)$ can be expressed as,
\begin{eqnarray}
H(t)\rightarrow {H_b(t)(1+\delta(t))}, \hspace{2cm} \rho(t)\rightarrow {\rho_b(t)(1+\delta_{m}(t))} \label{eq.67}
\end{eqnarray}

Both $\delta(t)$ and $\delta_{m}(t)$ can be seen as the isotropic deviation of the Hubble parameter and matter over-density. So, the perturbation of the function $f(Q)$ and $f_Q$ can be calculated as,
\begin{eqnarray}\label{eq.68}
\delta f=f_{Q}\delta Q, \hspace{2cm} \delta f_{Q}=f_{QQ}\delta Q, 
\end{eqnarray}

where $\delta Q$ represents the first order perturbation of the variable $Q$. Now, neglecting higher power of $\delta(t)$, the Hubble parameter can be obtained as,
\begin{equation}\label{eq.69}
6H^{2}=6H_b^{2}(1+\delta(t))^{2}=6H_b^{2}(1+2\delta(t))
\end{equation}
and subsequently eqn. \eqref{eq.10} can be reduced to
\begin{eqnarray} \label{eq.70}
Q(2Qf_{QQ}+f_{Q})\delta =\rho \delta_{m},
\end{eqnarray}

which gives the relation between the matter and geometric perturbation and the perturbed Hubble parameter can be realised from eqn. \eqref{eq.67}. Now, to obtain the analytical solution to the perturbation function, we consider the  perturbation continuity equation as, 
\begin{eqnarray}
\dot{\delta}_{m}+3H(1+\omega)\delta =0 \label{eq.71}
\end{eqnarray}
and from eqns. \eqref{eq.70}- \eqref{eq.71}, the following first order differential equation can be obtained, 
\begin{equation} \label{eq.72}
\dot{\delta}_{m}+\frac{3H(1+\omega)\rho}{Q(2Qf_{QQ}+f_{Q})}\delta_{m}=0   
\end{equation}

\begin{figure}[H]
\centering
\minipage{0.40\textwidth}
\includegraphics[width=\textwidth]{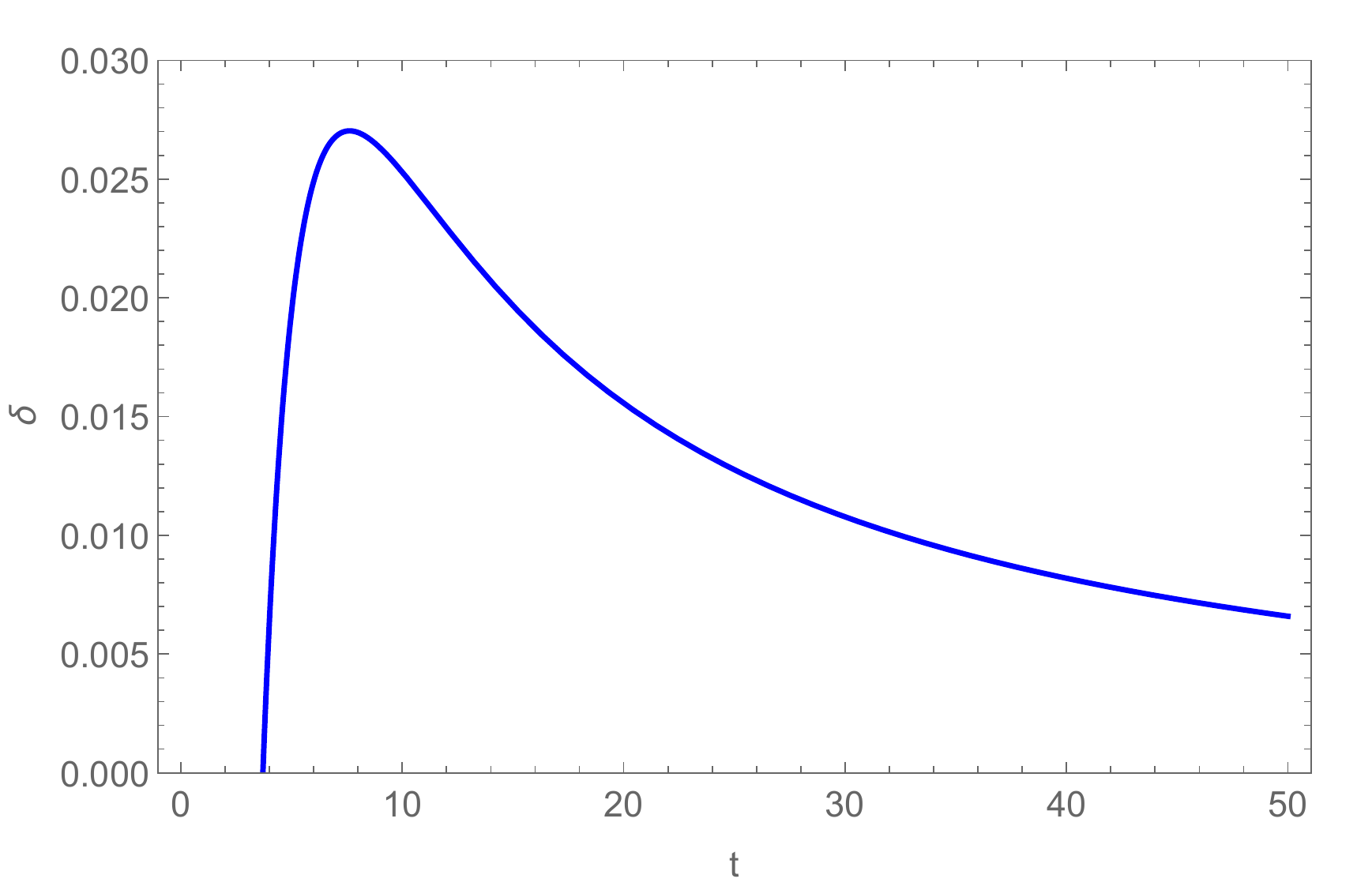}
\endminipage
\minipage{0.40\textwidth}
\includegraphics[width=\textwidth]{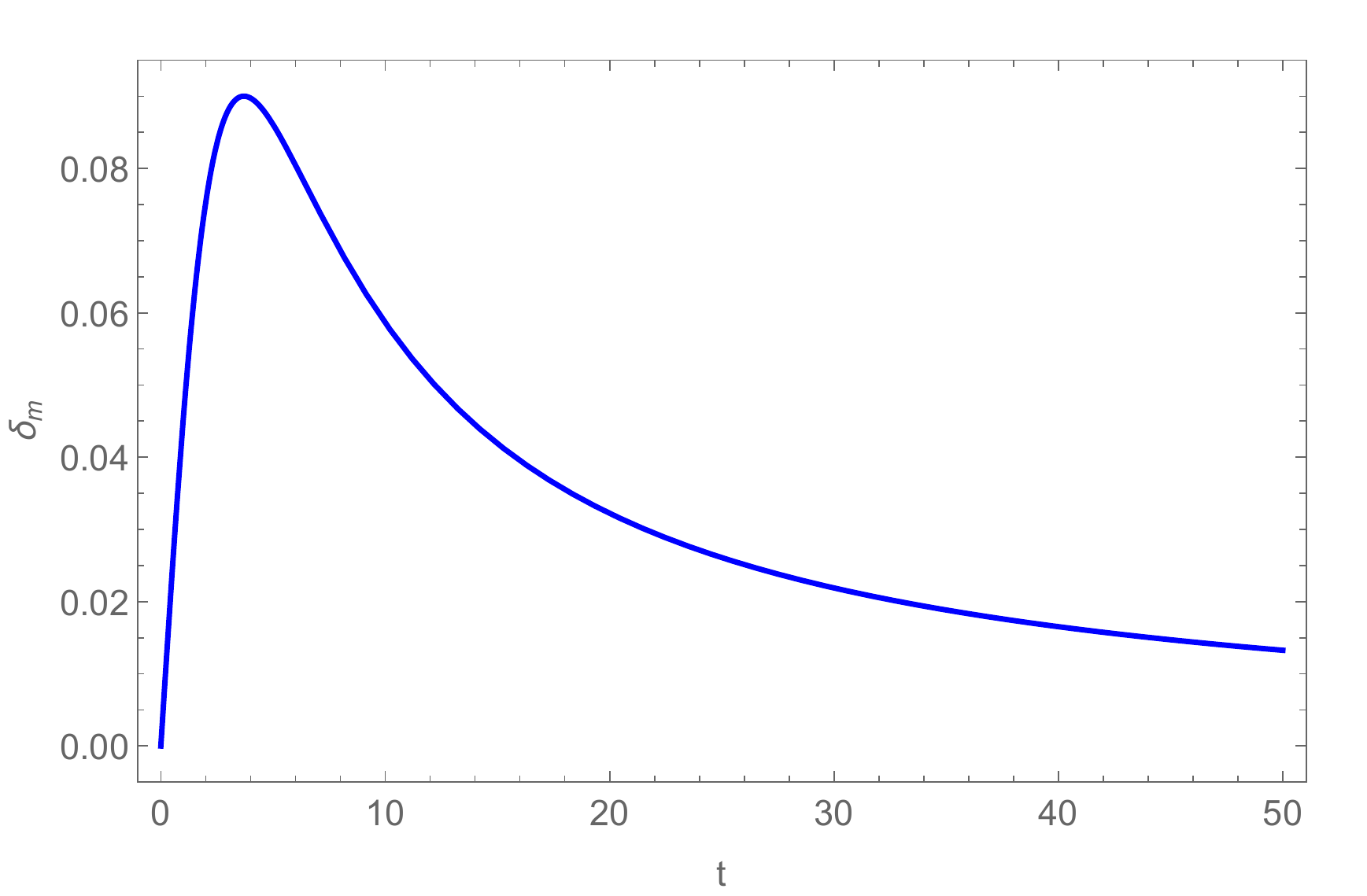}
\endminipage
\caption{Evolution of Hubble parameter and the energy density in cosmic time for $C_{1}=1$ and $C_{2}=-1/3$ (matter dominated case).}\label{FIG.4}
\end{figure}

Further using the $tt$-component field equation and eqn. \eqref{eq.72}, the simplified relation can be obtained,
\begin{equation}\label{eq.73}
\dot{\delta}_{m}-\frac{\dot{H}}{H}\delta_{m}=0 ,   
\end{equation}
which provides $\delta_{m}=C_{1}H$, where $C_{1}$ is the integration constant. Subsequently from eqn. \eqref{eq.71},  we obtain 
\begin{equation}\label{eq.74}
\delta=C_{2}\frac{\dot{H}}{H}    
\end{equation}
where, $C_{2}=-\frac{C_{1}}{3(1+\omega)}$. The evolution behaviour of $\delta$ and $\delta_m$ are given in FIG. \ref{FIG.4}. 

The pressure term becomes zero for the matter-dominated case, which implies that the equation of state parameter is also zero. So, with the equation of state parameter value zero, the stability of the model has been checked through the scalar perturbation of Hubble parameter and energy density. We can observe that at the start of both the deviations, $\delta(t)$ and $\delta_{m}(t)$, have some increment before declining through time and approaching zero at late times. As a result, we can say that though at the beginning the model shows unstable behaviour for a brief period, but in most of the time it shows stable behaviour under the scalar perturbation approach.

\section{Conclusion}\label{VII}

The matter bounce scenario of the Universe has been reconstructed in an extended symmetric teleparallel gravity. Considering the matter dominated phase at the background level, a specific form of $f(Q)$ has been developed that experiences the matter bounce scenario. The value of e-folding can also be obtained as a logarithmic function of an expression that contains the nonmetricity scalar. As expected, the model fails to explain the dark energy era, which has been observed from the dynamical stability analysis. Furthermore, with symmetric teleparallel gravity, the bottom-up reconstruction technique created a plausible non-singular bounce model. The bottom-up approach can be easily employed in an inflationary context when the observable quantities can be described in terms of the slow roll parameter in general (due to slow roll conditions). We employed a conformal equivalence between $f(Q)$ and scalar-tensor model to apply the bottom-up reconstruction technique in the $f(Q)$ bouncing model, because the slow roll requirement in the bouncing context is not true. In the scalar-tensor frame, the conformal factor is chosen in such a way that it results in an inflationary era. On the other hand a suitably considered conformal factor, the $f(Q)$ frame scale factor behaves as $a(\eta)=\left(\frac{3}{4}\rho_{c}\eta^{2}+1\right)^{1/3}$ which indicates a non-singular bounce at $\eta=0$.
From the critical points, the eigenvalues and the corresponding cosmology are obtained. Two critical points are obtained, one provides stable node and the other one unstable. The positive deceleration parameters show the decelerating Universe, occurred at early Universe. To check the stability of the reconstructed model, we have considered the scalar perturbation approach. From the graphical behavior of the deviation of the Hubble parameter and the energy density in cosmic time, it has been observed that both the deviations (i.e., $\delta(t)$ and $\delta_{m}(t)$) approaching zero at late times. We conclude that the reconstructed bouncing model though shows some amount of instability at the initial stage, but shows stability in most part of the evolution.

\section*{Acknowledgement} ASA acknowledges the financial support provided by University Grants Commission (UGC) through Senior Research Fellowship (File No. 16-9 (June 2017)/2018 (NET/CSIR)), to carry out the research work. BM acknowledge the support of IUCAA, Pune (India) through the visiting associateship program. The authors are thankful to the anonymous referee for the constructive suggestions and comments to improve the quality of the manuscript.


\begin{thebibliography}{99}
\section*{References}
\bibitem{Brout78} R. Brout, F. Englert, E. Gunzig, \href{https://doi.org/10.1016/0003-4916(78)90176-8}{\textit{Annals of Physics}, \textbf{115}, 78-106 (1978).}

\bibitem{Starobinsky80} A.A. Starobinsky, \href{https://doi.org/10.1016/0370-2693(80)90670-X}{\textit{Physics Letters B}, \textbf{91}, 99-102 (1980).}

\bibitem{Guth81} A.H. Guth, \href{https://doi.org/10.1103/PhysRevD.23.347}{\textit{Phys. Rev. D}, \textbf{23}, 347 (1981).}

\bibitem{Ashtekar06} A. Ashtekar, T. Pawlowski, P. Singh, \href{https://doi.org/10.1103/PhysRevD.74.084003}{\textit{Phys. Rev. D}, \textbf{74}, 084003 (2006).}

\bibitem{Sami06} M. Sami, P. Singh, S. Tsujikawa, \href{https://doi.org/10.1103/PhysRevD.74.043514}{\textit{Phys. Rev. D}, \textbf{74}, 043514 (2006).}

\bibitem{Ashtekar07} A. Ashtekar, \href{https://doi.org/10.1393/ncb/i2007-10351-5}{\textit{Nuovo Cim. B}, \textbf{122}, 135 (2007).}

\bibitem{Copeland08} E. J. Copeland, D. J. Mulryne, N. J. Nunes, M. Shaeri, \href{https://doi.org/10.1103/PhysRevD.77.023510}{\textit{Phys. Rev. D}, \textbf{77}, 023510 (2008).}

\bibitem{Corichi09} A. Corichi, P. Singh, \href{https://doi.org/10.1103/PhysRevD.80.044024}{\textit{Phys. Rev. D}, \textbf{80}, 044024 (2009).}

\bibitem{Bojowald09} M. Bojowald, \href{https://doi.org/10.1088/0264-9381/26/7/075020}{\textit{Class. Quant. Grav.} \textbf{26}, 075020 (2009).}

\bibitem{Ashtekar11} A. Ashtekar, P. Singh, \href{https://doi.org/10.1088/0264-9381/28/21/213001}{\textit{Class. Quant. Grav.}, \textbf{28}, 213001 (2011).}

\bibitem{Cai09} Y.F. Cai, T. Qiu, R. Brandenberger, X. Zhang, \href{https://doi.org/10.1103/PhysRevD.80.023511}{\textit{Phys. Rev. D}, \textbf{80}, 023511 (2009).}

\bibitem{Cai13} Y.F. Cai, E. McDonough, F. Duplessis, R. Brandenberger, \href{https://doi.org/10.1088/1475-7516/2013/10/024}{ \textit{JCAP}, \textbf{10} 024 (2013).} 
 
\bibitem{Quintin14} J. Quintin, Y.F. Cai, R.H. Brandenberger, \href{https://doi.org/10.1103/PhysRevD.90.063507}{\textit{Phys. Rev. D}, \textbf{90}, 063507 (2014).} 

\bibitem{Haro15} J. de Haro, Y.F. Cai, \href{https://doi.org/10.1007/s10714-015-1936-y}{ \textit{Gen. Rel. Grav.}, \textbf{47}, 95 (2015).}   

\bibitem{Saidov10} T. Saidov, A. Zhuk, \href{http://dx.doi.org/10.1103/PhysRevD.81.124002}{\textit{Phys. Rev. D}, \textbf{81}, 124002 (2010).}

\bibitem{Barragan10} C. Barragan, G.J. Olmo, \href{https://doi.org/10.1103/PhysRevD.82.084015}{\textit{Phys. Rev. D}, \textbf{82}, 084015 (2010).}

\bibitem{Cai11} Y.F. Cai, S.H. Chen, J.B. Dent, S. Dutta, E.N. Saridakis, \href{https://doi.org/10.1088/0264-9381/28/21/215011}{\textit{Class. Quantum Grav.}, \textbf{28}, 215011 (2011).}

\bibitem{Battefeld15} D. Battefeld, P. Peter, \href{https://doi.org/10.1016/j.physrep.2014.12.004}{\textit{Phys Reports}, \textbf{571}, 1 (2015).}

\bibitem{Odintsov15} S.D. Odintsov, V.K. Oikonomou, E.N. Saridakis, \href{https://doi.org/10.1016/j.aop.2015.08.021}{\textit{Ann. Phys.}, \textbf{363}, 141 (2015).}

\bibitem{Brandenberger17} R. Brandenberger, P. Peter, \href{https://doi.org/10.1007/s10701-016-0057-0}{\textit{Found. Phys.}, \textbf{47}, 797 (2017).}

\bibitem{Hohmann17} M. Hohmann, L. Jarv, U. Ualikhanova, \href{http://dx.doi.org/10.1103/PhysRevD.96.043508}{\textit{Phys. Rev. D}, \textbf{96}, 043508 (2017).}

\bibitem{Ijjas18} A. Ijjas, P.J.  Steinhardt, \href{https://doi.org/10.1088/1361-6382/aac482}{\textit{Class. Quantum Grav.}, \textbf{35}, 135004 (2018).}

\bibitem{Dombriz18}  A. de la Cruz-Dombriz, G. Farrugia, J. Levi Said, D.S.C. Gomez, \href{http://dx.doi.org/10.1103/PhysRevD.97.104040}{\textit{Phys. Rev. D}, \textbf{97}, 104040 (2018).}

\bibitem{Shabani18} H. Shabani, A. H. Ziaie, \href{https://doi.org/10.1140/epjc/s10052-018-5886-x}{\textit{Eur. Phys. J. C}, \textbf{78}, 397 (2018).}

\bibitem{Matsui19} H. Matsuia, F. Takahashi, T. Terada, \href{https://doi.org/10.1016/j.physletb.2019.06.013}{\textit{Phys. Lett. B}, \textbf{795}, 152 (2019).}

\bibitem{Caruana20} M. Caruana, G. Farrugi, J. Levi Said, \href{https://doi.org/10.1140/epjc/s10052-020-8204-3}{\textit{Eur. Phys. J. C}, \textbf{80}, 640 (2020).} 

\bibitem{Mishra21} B. Mishra, F. Md. Esmeili, S. Ray, \href{https://doi.org/10.1007/s12648-020-01877-2}{\textit{Indian J. Phys.} \textbf{95}, 2245 (2021).}

\bibitem{Tripathy21} S.K. Tripathy, B.Mishra, S. Ray, R. Sengupta, \href{https://doi.org/10.1016/j.cjph.2021.03.026}{\textit{Chinese J. Phys.}, \textbf{71}, 610 (2021).}

\bibitem{Agrawal21a} A.S. Agrawal, F. Tello-Ortiz, B. Mishra, S.K. Tripathy, \href{https://doi.org/10.1002/prop.202100065}{\textit{Fortschritte der Physik}, \textbf{70}, 2100065 (2021).}

\bibitem{Agrawal22} A.S. Agrawal,  S.K. Tripathy, S. Pal, B. Mishra, \href{https://doi.org/10.1088/1402-4896/ac49b2}{\textit{Phys. Scr.}, \textbf{97}, 025002  (2022).}

\bibitem{Odintsov14} S.D. Odintsov, V.K. Oikonomou, \href{https://doi.org/10.1103/PhysRevD.90.124083}{\textit{Phys. Rev. D}, \textbf{90}, 124083 (2014).}

\bibitem{Odintsov20} S.D.Odintsov, V.K. Oikonomou, T. Paul,  \href{https://doi.org/10.1016/j.nuclphysb.2020.115159}{\textit{Nuclear Physics B}, \textbf{959}, 115159 (2020).}
 
\bibitem{Odintsov21} S.D. Odintsov, T. Paul, I. Banerjee, R. Myrzakulov, S. SenGuptah, \href{https://doi.org/10.1016/j.dark.2021.100864}{\textit{Phys. Dark Univ.}, \textbf{33}, 100864 (2021).} 

\bibitem{Belinskii70} V.A. Belinskii, I.M. Khalatnikov, E.M. Lifshitz, \href{https://doi.org/10.1080/00018737000101171}{\textit{Advances in Physics}, \textbf{19}, 525 (1970).}

\bibitem{Jimenez18} J. B. Jimenez, L. Heisenberg, T. Koivisto, \href{https://doi.org/10.1103/PhysRevD.98.044048}{\textit{Phys. Rev. D} \textbf{98}, 044048 (2018).}

\bibitem{Lazkoz19} R. Lazkoz, F.S.N. Lobo, M. Ortiz-Banos, V. Salzano, \href{https://doi.org/10.1103/PhysRevD.100.104027}{\textit{Phys. Rev. D} \textbf{100}, 104027 (2019).}

\bibitem{Bajardi20} F. Bajardi, D. Vernieri, S. Capozziello, \href{https://doi.org/10.1140/epjp/s13360-020-00918-3}{\textit{Eur. Phys. J. Plus} \textbf{135}, 912 (2020).} 

\bibitem{Agrawal21b} A.S. Agrawal, L. Pati, S.K. Tripathy, B. Mishra, \href{https://doi.org/10.1016/j.dark.2021.100863}{\textit{Phys. Dark Univ.}, \textbf{33}, 100863 (2021).}

\bibitem{Narawade22} S.A. Narawade, L. Pati, B. Mishra, S.K. Tripathy, \href{https://doi.org/10.1016/j.dark.2022.101020}{\textit{Phys. Dark Univ.}, \textbf{36},  101020 (2022)}

\bibitem{Jarv18} L. Jarv, M. Runkla, M. Saal, O. Vilson, \href{https://doi.org/10.1103/PhysRevD.97.124025}{\textit{Phys. Rev. D}, \textbf{97},
124025 (2018).}

\bibitem{Hehl95} F. W. Hehl, J. D. McCrea, E. W. Mielke, Y. Neeman, \href{https://doi.org/10.1016/0370-1573(94)00111-F}{\textit{Phys. Rep.} 258, 1 (1995).}

\bibitem{Ortin15} T. Ortin, \href{https://doi.org/10.1017/CBO9781139019750}{Gravity and Strings (Cambridge University Press, Cambridge, England, 2015).}

\bibitem{Jimenez20} J.B. Jimenez, L. Heisenberg, T. Koivisto, S. Pekar, \href{https://doi.org/10.1103/PhysRevD.101.103507}{\textit{Phys. Rev. D}, \textbf{101}, 103507 (2020).}

\bibitem{Haro14} J. Haro, J. Amoros,    \href{https://doi.org/10.1088/1475-7516/2014/12/031}{\textit{JCAP} \textbf{12} 031 (2014).}

\bibitem{Miranda22} M. Miranda, D. Vernieri, S. Capozziello, F.S.N. Lobo, \href{https://doi.org/10.3390/universe8030161}{\textit{Universe}, \textbf{8(3)}, 161 (2022).}

\bibitem{Nojiri09} S. Nojiri, S.D. Odintsov, D. Saez-Gomez
\href{https://doi.org/10.1016/j.physletb.2009.09.045}{\textit{Phys. Lett. B}, \textbf{681}, 74 (2009).}

\bibitem{Amoros13} J. Amoros, J. de Haro, S. D. Odintsov, \href{https://doi.org/10.1103/PhysRevD.87.104037}{\textit{Phys. Rev. D}, \textbf{87}, 104037 (2013).}

\bibitem{Nojiri17} S.Nojiri, S.D.Odintsov, V.K.Oikonomou, \href{https://doi.org/10.1016/j.physrep.2017.06.001}{\textit{Phys. Rep.} \textbf{692} 1-104 (2017)}

\bibitem{Hwang05} J.c. Hwang, H. Noh, \href{https://doi.org/10.1103/PhysRevD.71.063536}{\textit{Phys. Rev. D} \textbf{71}, 063536 (2005).}

\bibitem{Khyllep21} W. Khyllep, A. Paliathanasis, J. Dutta, \href{https://doi.org/10.1103/PhysRevD.103.103521}{\textit{Phys. Rev. D}, \textbf{103}, 103521 (2021).}

\bibitem{Wu12} Y.-P. Wu, C.-Q. Geng, \href{https://doi.org/10.1007/JHEP11(2012)142}{\textit{J. High Energ. Phys.} \textbf{2012}, 142 (2012).}

\bibitem{Golovnev18} A. Golovnev, T. Koivisto, \href{https://doi.org/10.1088/1475-7516/2018/11/012}{\textit{JCAP} \textbf{11}, 012 (2018).}

\bibitem{Duchaniya22} L. K. Duchaniya, S. V. Lohakare, B. Mishra, S. K. Tripathy, \href{https://doi.org/10.1140/epjc/s10052-022-10406-w}{\textit{Eur. Phys. J. C}, \textbf{82}, 448 (2022).}

\bibitem{Akrami20} Y. Akrami \href{https://doi.org/10.1051/0004-6361/201833887}{\textit{Aston. \& Astrophy.}, \textbf{641}, A10 61  2020.}


%%%%%%%%%%%%%%%%%%%%%%
\end{thebibliography}
\end{document}